\documentclass[showpacs,twocolumn,superscriptaddress]{revtex4}
\usepackage{epsfig}
\usepackage{amssymb}
\usepackage{amsmath}
\usepackage{graphicx}

\usepackage{epsfig}
\usepackage{graphicx}
\usepackage{amsmath,amsfonts,amssymb}
\usepackage{pstricks}
\usepackage{color}
\usepackage{soul}

\begin{document}

\title{The critical Ising model on a torus with a defect line}

\date{\today}

\author{Armen Poghosyan}
\affiliation{Yerevan Physics Institute,
Alikhanian Brothers 2, 375036 Yerevan, Armenia}
\author{Ralph Kenna}
\affiliation{Applied Mathematics
Research Centre, Coventry University, Coventry CV1 5FB, UK}
\author{Nikolay Izmailian}
\affiliation{Yerevan Physics Institute,
Alikhanian Brothers 2, 375036 Yerevan, Armenia}
\affiliation{Applied Mathematics Research Centre, Coventry University, Coventry CV1 5FB, UK}

\begin{abstract}
The critical Ising model in two dimensions with a defect line is analyzed to deliver the first exact solution with twisted boundary conditions. We derive exact expressions for the eigenvalues of the transfer matrix and obtain analytically the partition function and the asymptotic expansions of the free energy and  inverse correlation lengths for an infinitely long cylinder of circumference $L_x$.
We find that finite-size corrections to scaling are of the form $a_k/L^{2k-1}_x$ for the free energy $f$ and $b_k(p)/L_x^{2k-1}$ and $c_k(p)/L_x^{2k-1}$ for inverse correlation lengths $\xi^{-1}_p$ and $\xi^{-1}_{L-p}$, respectively, with integer values of $k$.
By exact evaluation we find that the amplitude ratios $b_k(p)/a_k$ and $c_k(p)/a_k$ are universal and verify this universal behavior using a perturbative conformal  approach.
\end{abstract}

\maketitle

The study of boundary conditions (BC's) in conformal field theories has attracted much attention in recent decades.
This is because of its relevance in string and brane theory on the one hand, and in various problems of statistical mechanics and condensed matter physics on the other.
A  classification of  conformal boundary conditions on the torus has been given in \cite{Affleck1997,Petkova,Chui}.
Computation of the partition function involves identifying the states at the two ends of a cylinder through the trace operation. One may insert a defect line (or seam) into such a system, along a non-contractible circle at the end of the cylinder before closing it into a torus, the effect of which is to twist the boundary conditions. It is important to understand the effects of such a twist, and to do this it is valuable to study model systems, especially those which have exact results.
The Ising model is the most prominent example and one of the best studied models of statistical mechanics.
Since Onsager obtained the exact solution of the two-dimensional Ising model with cylindrical BC's in 1944 \cite{Onsager}, there have been continuous attempts to treat different two-dimensional topologies \cite{Kaufman,LuWu2001,LuWu1998,Liaw,Brascamp,Brien}.

At the critical point $T_c$ the asymptotic finite-size scaling
behavior of the critical free energy $f_N$ and the inverse
correlation lengths $\xi_{{n}}^{-1}$  of an infinitely long 2D cylinder
of finite circumference $N$ has the form \cite{Blote}
\begin{equation}
\lim_{N \to \infty}\left[ N^2 (f_N-f_{\infty})\right]=A, \quad \lim_{N \to \infty} N\xi_n^{-1}=D_n \nonumber
\end{equation}
where $f_{\infty}$ is the bulk free energy and A and $D_n$ are the universal constants which may depend on the BC's. The values of A and $D_n$ are known  to be related to the conformal anomaly  $c$, the conformal weight $\Delta$, and the scaling dimension
of the $n$-th scaling field $x_n$ of the theory \cite{Blote}
\begin{equation}
A=2\pi \zeta \left(-\frac{c}{12}+\Delta+\bar \Delta\right), \qquad
D_n=2\pi \zeta x_n,
\nonumber
\end{equation}
where $\zeta$ is anisotropy parameter. The
principle of unitarity of the underlying field theory restricts,
through the Kac formula, the possible values of $c$, $\Delta$ and $\bar \Delta$.
For the 2D Ising model, we have $c = 1/2$
and the only possible values are $\Delta, \bar \Delta = 0, 1/16, 1/2$. Therefore there are six different boundary universal classes with $(\Delta, \bar \Delta) = (0,0)$  for periodic
BC's; $(\Delta, \bar \Delta) = \left(\frac{1}{16}, \frac{1}{16}\right)$ for antiperiodic BC's; and BC's with $(\Delta, \bar \Delta) = \left( \frac{1}{2}, \frac{1}{2}\right); \left(\frac{1}{16},0\right); \left(\frac{1}{2},0\right)$ and $\left( \frac{1}{16}, \frac{1}{2}\right)$. Past efforts have focused mainly on periodic and antiperiodic BC's. In this paper we consider $(\Delta, \bar \Delta)= \left(\frac{1}{16}, 0\right)$, which for the Ising quantum chain is called duality twisted BC's and has been considered in \cite{Schutz}. In what follows we will give the  first exact solution of the 2D Ising model with twisted BC's at the critical point. New universal amplitude ratios for finite-size corrections of the two-dimensional Ising model with periodic, antiperiodic, free, fixed and mixed boundary conditions have been recently presented
\cite{Izmailian2001}. In this letter we  present universal amplitude ratios for the duality twisted BC's universality class.

The Ising model on a lattice with periodic BC's and with a specific defect line (seam) was formulated in \cite{Brien,Pearce2001}. For the Ising model the seams are labelled by the Kac labels $(r,s)$.  There are six possible partition functions $Z_{(r,s)}(q)$ for the Ising model with seams labelled by $(r,s)= (1,1), (1,2), (1,3), (2,1), (2,2), (2,3)$ and for the three of them, namely, for the seams $(r,s)= (1,1), (1,2)$ and $(1,3)$, the partition functions $Z_{(r,s)}(q)$ are obtained numerically to very high precision in \cite{Pearce2001}. In particular for $Z_{(1,2)}(q)$ they obtain
\begin{eqnarray}
Z_{(1,2)}(q) &=&  \left[\chi_0+\chi_{\frac{1}{2}}\right]\chi_{\frac{1}{16}}^*+
\left[\chi_0+\chi_{\frac{1}{2}}\right]^*\chi_{\frac{1}{16}}, \label{Z12}
\end{eqnarray}
where $\chi_0 \equiv \chi_0(q)$, $\chi_{\frac{1}{2}} \equiv \chi_{\frac{1}{2}}(q)$ and $\chi_{\frac{1}{16}} \equiv \chi_{\frac{1}{16}}(q)$ are the chiral characters and $q$ is the modular parameter. The $(r, s) = (1, 1)$ and $(1, 3)$ seams reproduce the well-known partition function of the Ising model with  periodic and antiperiodic boundary conditions respectively. In what follows we will show that the BC given by the seam $(r, s) = (1, 2)$ corresponds to duality twisted BC.
\begin{figure}[t]
\begin{center}
\includegraphics[height=4.778cm,width=7cm,angle=0]{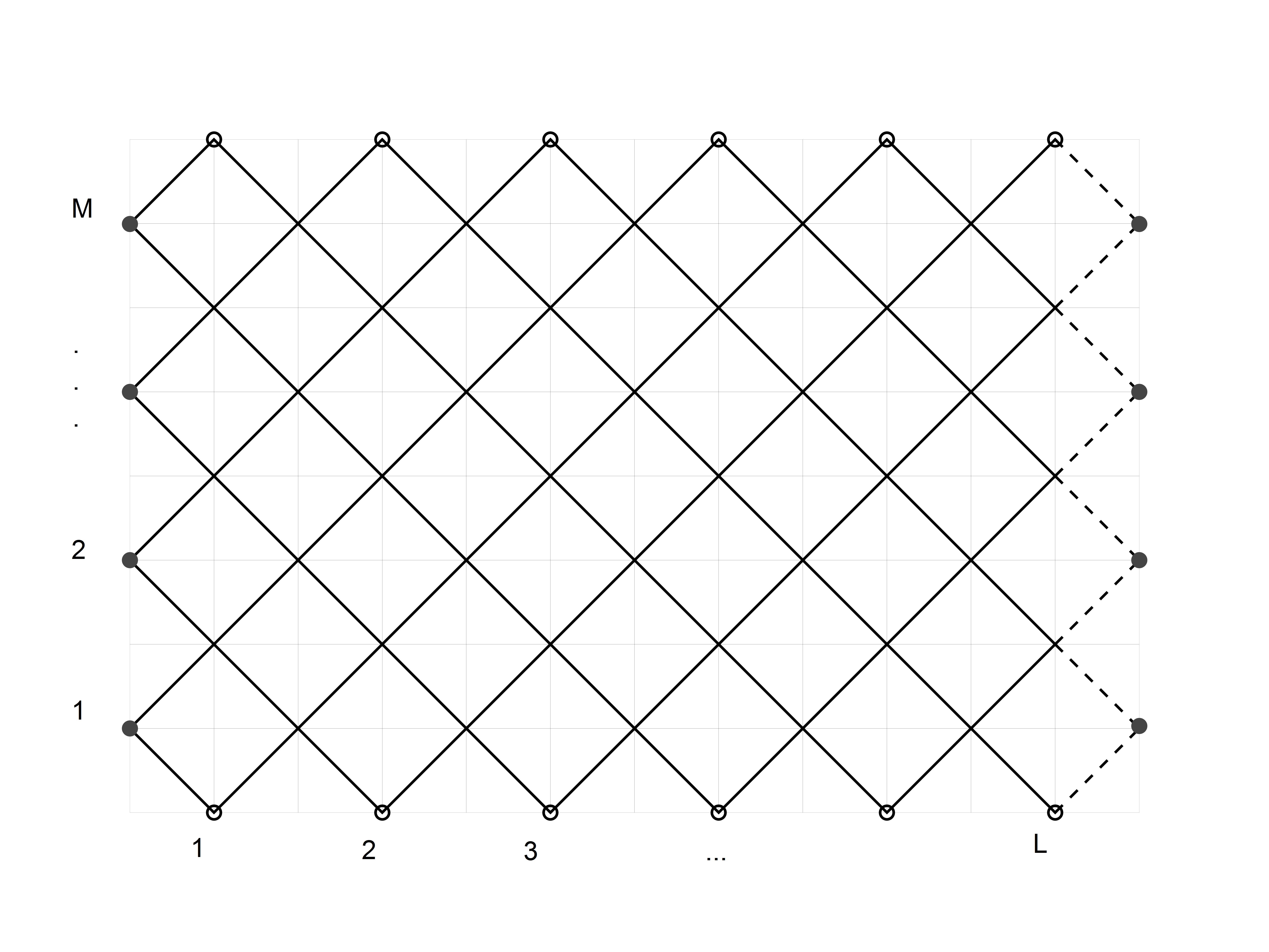}
\end{center}
\caption{The lattice is constructed with $L$ faces in each row and $M$ faces in each column.
The line defect is inserted along the rightmost columnar edge (dashed line) before closing the lattice into a torus with periodic BC's.}
\end{figure}

Let us consider a square lattice rotated by $45$ degrees (Fig. 1), in which each row has $L$ faces
and each column has $M$ faces.
We insert one defect line (seam) along the final column.
The lattice thus consists of $2L-1$ regular (zigzagging) columnar edges and one defect line and $2M$ regular (zigzagging) row edges.
Periodic BC's are imposed in both directions and in Fig. 1 this is represented as identifying  the light (dark) nodes on the first row (column) with the respective light (dark) nodes on the last row (column). The physical dimensions of the lattice, $L_x$ and $L_y$, are given by
\begin{equation}
L_x=\sqrt{2}\left(L-\frac{1}{2}\right); \qquad L_y=\sqrt{2}\;M. \label{Lx}
\end{equation}
The finite-size partition function for the Ising model $Z_{L_x L_y}$ can be written as $Z_{L_x L_y} = \sum_{s}\exp{\left(J\sum_{<ij>} s_i s_j+K\sum_{<ij>}s_i s_j\right)}$,
where the first sum within the parenthesis is over NW-SE edges, and the second sum over NE-SW edges and spin variable $s_i$ can take the two values $\pm 1$. Since we restrict ourselves to the critical Ising model, we have $\sinh(2J) \sinh(2K) = 1$. This condition can be conveniently parameterized by introducing a so-called spectral parameter $u$, so that $\sinh(2J) = \cot(2u), \sinh(2K) = \tan(2u)$, with $0<u<\pi/4$. The anisotropy parameter is $\zeta$ related to the spectral parameter $u$ through $\zeta = \sin 4u$.
Now partition function can be rewritten in the following form
\begin{equation}
Z_{L_x L_y}(u) = Tr\left[T(u)\right]^M
=\sum_{n}e^{-L_y {\cal E}_n(u)}
\nonumber
\end{equation}
where the sum is over all eigenvalues of a transfer matrix $T(u)$, written as $e^{-{\bf E}_n(u)}$ and ${\cal E}_n(u)=\frac{\sqrt{2}}{2} {\bf E}_n(u)$.

Conformal invariance predicts that the leading finite-size corrections to the energies ${\cal E}_n$ take the form \cite{Chui}
\begin{eqnarray}
{\cal E}_n(u)&=&L_x f_{\infty}+\frac{2\pi i}{L_x}\left[\left(\Delta_n+\bar k_n - \frac{c}{24}\right)e^{-i g u} \right. \nonumber \\
&-& \left. (\bar\Delta_n+\bar k_n-\frac{c}{24})e^{i g u}\right]+O\left(\frac{1}{L_x}\right),
\label{energy}
\end{eqnarray}
where $\Delta_n$ and $\bar\Delta_n$ are the conformal weights, $k_n, \bar k_n \in N$ are label descendent levels and $g$ is Coxeter number.

The Boltzmann weights of the model are prescribed to the faces of a regular square lattice \cite{Pearce2001}
\begin{eqnarray}
\label{BW}
W\left(\begin{array}{cc|}a & b \\ c & d \end{array}\;u\right)
=s_1(-u)\delta_{ac}+
s_0(u)\frac{\sqrt{\psi_a\psi_c}}{\psi_b}\delta_{bd}.
\end{eqnarray}
Here $a,b,c,d$ are the spin states, $u$ ($0<u<\lambda$) is the spectral parameter, $s_k(u)=\frac{\sin(u+k\lambda)}{\sin\lambda}$ , $\lambda=\frac{\pi}{g}$ is the crossing parameter and $\psi_a$ are the entries of Perron-Frobenius eigenvector of the adjacency matrix $G$. The Ising model is related to the Dynkin diagram $A_{3}$ (see Fig. 2) whose Coxeter number $g=4$.
\begin{figure}[t]
\begin{center}
\includegraphics[height=0.865cm,width=7cm,angle=0]{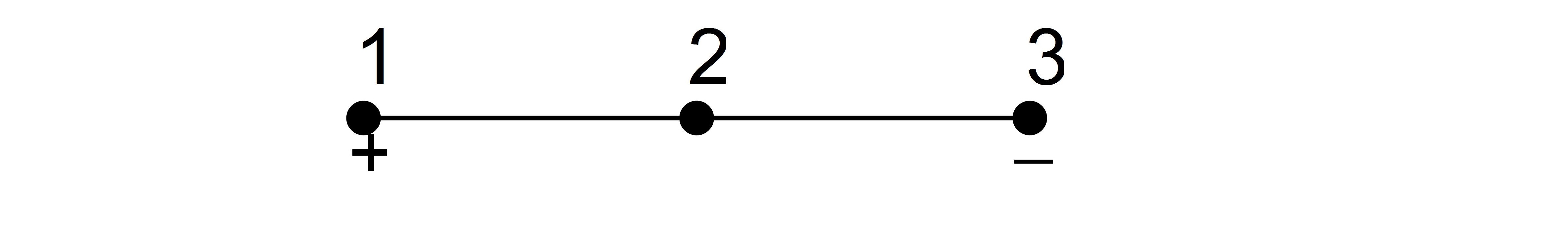}
\caption{$A_3$ Dynkin diagram}
\end{center}
\end{figure}
From $A_{3}$ Dynkin diagram we get  the following allowed face configurations:
\begin{figure}[htb]
\label{F}
\begin{center}
\includegraphics[height=1.352cm,width=7cm,angle=0]{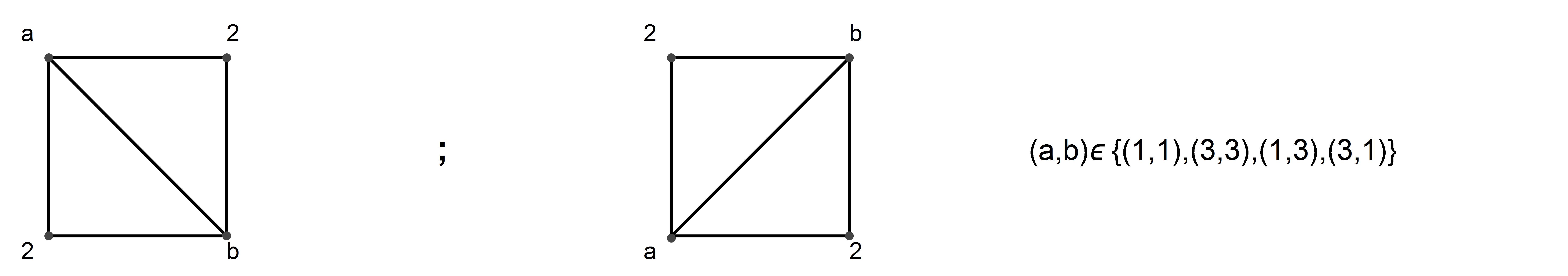}
\end{center}
\end{figure}

We can calculate the weights of this faces by inserting the respective parameters in equation (\ref{BW}).
Obviously one can construct two kinds of transfer matrices $U$ and $V$ (see Fig. 3)
\begin{figure}[htb]
\begin{center}
\includegraphics[height=5cm,width=7cm,angle=0]{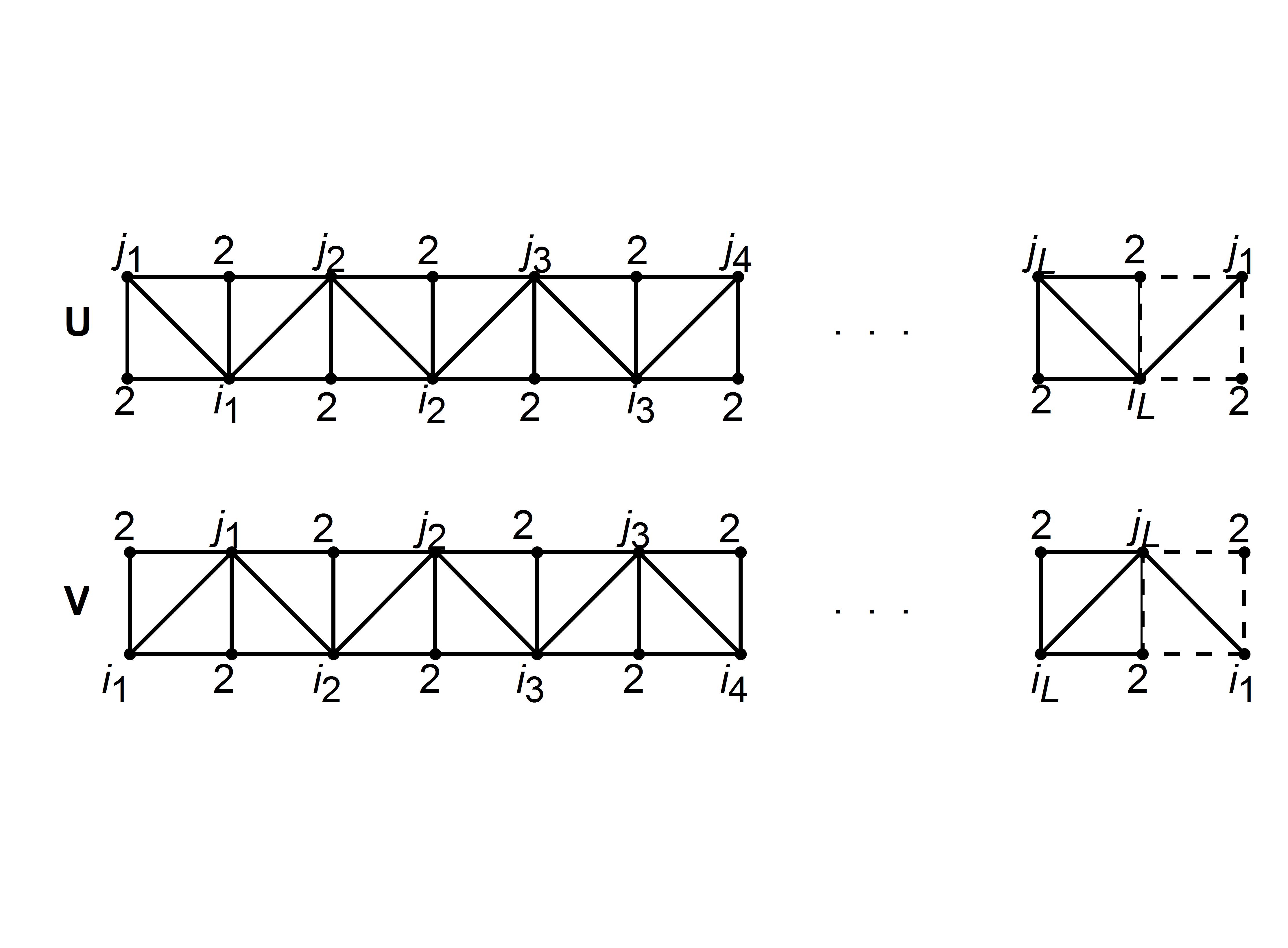}
\end{center}
\caption{$U$ and $V$ transfer matrices. The rows of the transfer matrices consist of $2L-1$ regular faces followed by single seams. Single seams represents by dashed lines.}
\end{figure}
where the rows consist of $2L-1$ regular faces followed by single seams. Our BC are given with seam $(r,s)=(1,2)$. The indices $i_{k}$, $j_{k}$ take values $1$ or $3$.
In terms of these quantities we can construct the double row transfer matrix $U(u)V(v)$ (see Fig. 4).
\begin{figure}[t]
\begin{center}
\includegraphics[height=5cm,width=7cm,angle=0]{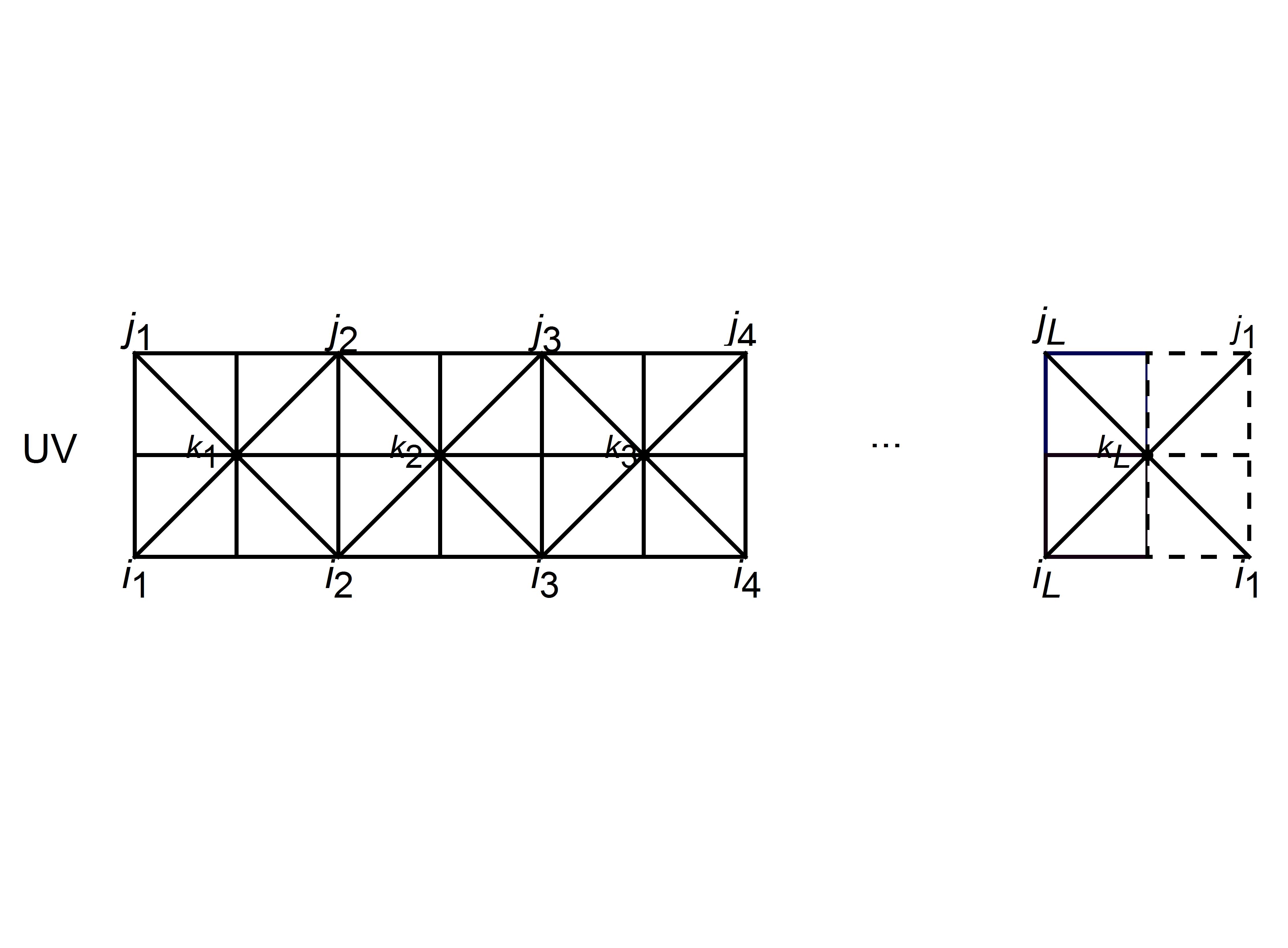}
\end{center}
\caption{The $UV$ double row transfer matrix consists of $L-1$ blocks and the last dashed block.}
\end{figure}
Where $i_a,j_a,k_a$ take values $1$ or $3$ ($a=1,...,L$), and over the variables $k_1,k_2,...,k_L$ a summation is performed. On other vertexes assume non-fluctuating heights equal to $2$. The double row transfer matrix $U(u)V(v)$ satisfies the functional equation
\begin{eqnarray}
U(u)V\left(u+\frac{\pi}{4}\right)=\cos\left(2u\right)^{2L-1}I+i\left(-1\right)^{L}\sin\left(2u\right)^{2L-1}R \nonumber
\end{eqnarray}
where  $I$ is the identity matrix, and R is a square matrix with anti-diagonal
entries equal to 1 with remaining entries $0$ [both are ($2^{L}\times 2^{L}$) matrices]. For the eigenvalues $\Lambda(u)$ of the double row transfer matrices $UV$  of the Ising model with duality twisted BC's we can obtain
\begin{eqnarray}
\Lambda=e^{2 i \bar u} \prod _{k=1}^{L-1} \left(e^{2 i \bar u} \sin \varphi_k+\mu _k e^{-2 i \bar u} \cos \varphi_k \right)^2 \label{l new}
\end{eqnarray}
where $\bar u = u-\frac{\pi }{8}$, $\varphi_k= \frac{\pi(2k-1)}{2(2L-1)}$ and $\mu_k=\pm 1$ can be chosen arbitrarily so we have $2^{L-1}$ eigenvalues. The remaining eigenvalues can be found by taking the complex conjugate of (\ref{l new}).
From Eq. (\ref{l new}) one can easily obtain the expressions for the absolute value of the eigenvalue $|\Lambda|$ and the argument $\theta$.
Let us now consider the largest eigenvalue $\Lambda_0$, which corresponds to the case when all $\mu_k=1$. The derivation of the asymptotic expansion of $\log \Lambda_0 = \log \left|\Lambda_0\right| + i \theta_0$ can be divided into two parts. First,  with the help of the Euler-Maclaurin summation formula we can derive the asymptotic expansion of the logarithm of the absolute value of $\Lambda_0$
\begin{eqnarray}
\frac{\sqrt{2}}{2}\log \left|\Lambda_0\right|&=& -L_x f_{\infty}-\sum_{k=0}^{\infty}\frac{B_{2k+2}f^{(2k+1)}(0)}{2^k(2k+2)!}\left(\frac{\pi}{L_x}\right)^{2k+1}
\nonumber
\end{eqnarray}
where $B_{2k+2}$ are the Bernoulli numbers, $f(x)=\frac{1}{2}\log\left[1+\sin(4u) \sin x \right]$ and $f_{\infty}=-\frac{1}{\pi}\int_{0}^{\pi} f(x) dx$.
Next, with the help of the Boole summation formula \cite{Borwein} the asymptotic expansion of the argument $\theta_0$ can be written in the  form
\begin{eqnarray}
\theta_0&=&-\frac{\sqrt{2}}{2}\sum_{n=0}^{\infty}\frac{E_{2n+1}(0)g^{(2n+1)}(0)}{2^n(2n+1)!}
\left(\frac{\pi}{L_x}\right)^{2n+1},
\nonumber
\end{eqnarray}
where $g(x)=\frac{1}{2}\arctan{\frac{\cos(4u)\cos x}{\sin(4u)+\sin x}}$ and $E_{2n+1}(0)$ are the Euler polynomials.
Thus the leading finite-size corrections to the ground state energies ${\cal E}_0=\frac{\sqrt{2}}{2} {\bf E}_0 =-\frac{\sqrt{2}}{2}\log \Lambda_0$ take the form
\begin{eqnarray}
\label{log l_0}
{\cal E}_0&=& L_x f_{\infty}+\frac{2\pi i}{L_x}
\left(\frac{1}{24}e^{-4 i u}
+\frac{1}{48}e^{4 i u}\right).\label{E0Ising}
\end{eqnarray}
On the other hand, ${\cal E}_0$ given by Eq. (\ref{energy})
Comparing Eq. (\ref{E0Ising}) with Eq. (\ref{energy}) for the Ising model with $c=1/2$ and $g=4$  we  see that $\Delta = \frac{1}{16}$ and $\bar \Delta = 0$. Thus we have shown that duality twisted BC's with $\left(\Delta, \bar \Delta\right) = \left(\frac{1}{16}, 0\right)$ correspond to the BC with seam $(r,s)=(1,2)$.

Let us now consider the other eigenvalues given by Eq. (\ref{l new}), which  correspond to some combination of the $\mu_k$.
Denote by $\Lambda_{\{p\}}$ the eigenvalue  with $\mu_p=-1$ and the remaining $\mu_k$ having the value $\mu_k=+1$.
The leading finite-size corrections to the exited state energies ${\cal E}_{p}=\frac{\sqrt{2}}{2} {\bf E}_{p} =-\frac{\sqrt{2}}{2}\log \Lambda_{\{p\}}$ and ${\cal E}_{L-p}=\frac{\sqrt{2}}{2} {\bf E}_{L-p} =-\frac{\sqrt{2}}{2}\log \Lambda_{\{L-p\}}$ can be written in the  form
\begin{eqnarray}
{\cal E}_{p}&=&L_x f_{\infty}+\frac{2\pi i}{L_x}
\left[\frac{1}{24} e^{-4 i u}
+ \left(\frac{25}{48}-p\right)e^{4 i u}
\right]\label{Epf}\\
{\cal E}_{L-p}&=&L_x f_{\infty}+\frac{2\pi i}{L_x}
\left[\left(\frac{1}{24}+p\right) e^{-4 i u}
+ \frac{1}{48}e^{4 i u}
\right], ~\label{ELpf}
\end{eqnarray}
where $L_x$ is given by Eq. (\ref{Lx}).
Comparing Eqs. (\ref{Epf}), (\ref{ELpf}) with Eq. (\ref{energy}) for the Ising model with $c=1/2$ and $g=4$ we can see that for the exited state ${\cal E}_{p}$ we have $\Delta = 1/16$, $k_n=0$ and $\bar \Delta = 1/2$, $\bar k_n = p-1$ and for the exited state ${\cal E}_{L-p}$ we have $\Delta = 1/16$, $k_n=p$ and $\bar \Delta = 0$, $\bar k_n = 0$.

Now we have all the necessary information to start the calculations of partition
function for the Ising model with duality twisted BC's $Z_{L_xL_y}$. For large $L_x$ and $L_y$ (always keeping the ratio $L_y/L_x$ constant) we have
\begin{eqnarray}
Z_{L_x L_y}=\sum_l\Lambda_l^M\approx e^{-L_xL_yf_{\infty}}Z_{(1,2)}(q),
\nonumber
\end{eqnarray}
where $Z_{(1,2)}(q)$ is the universal conformal partition function. Let us now find the general form of the eigenvalues with significant input in the partition function. From (\ref{l new}) we can see that these are eigenvalues
for which almost all $\mu_k=1$ and some $\mu_k$ are allowed to take the value $-1$
only if $k\ll L$ or $L-k\ll L$. Any "significant" eigenvalue will be specified by two sets of indexes $K=\{k_1,k_2,\cdots,k_m\}$ and
 $\bar K=\{\bar k_1,\bar k_2,\cdots,\bar k_{\bar m}\}$, where $\bar k_i \equiv L-k_i$ with
 $k_1< k_2<k_3<\cdots<k_m\ll L $ and $\bar{ k_1}<\bar{k_2}<\bar{k_3}<\cdots
<\bar{k}_{\bar{ m}}\ll L$ so that $\mu_{\bar k_i}=-1$,$\mu_{k_i}=-1$ and the other
$\mu$'s are $+1$. Thus from above it is easy to get
\begin{eqnarray}
\label{UnPart}
Z_{(1,2)}(q)=\sum_{K\bar K}q^{-\frac{c}{24}+\frac{1}{16}+\sum_i k_{i}}
\bar{q}^{-\frac{c}{24}+\sum_i(\bar{k}_{i}-\frac{1}{2})}
\end{eqnarray}
where $q=e^{-\frac{2 \pi i L_y}{L_x} e^{-4 iu}}$ is a modular parameter.
For further calculations it is convenient to introduce the occupation numbers
$\varepsilon_k=\frac{1-\mu_k}{2}$ and $\bar{\varepsilon}_{\bar{k}}=\frac{1-\bar\mu_{\bar k}}{2}$.
In terms of these quantities the universal conformal partition function (\ref{UnPart}) can be
rewritten as
\begin{eqnarray}
&&Z_{(1,2)}(q)=\sum_{\{\varepsilon\}\{\bar\varepsilon\}}q^{-\frac{c}{24}+\frac{1}{16}+\sum_k^\infty k \varepsilon_{k}}
\bar{q}^{-\frac{c}{24}+\sum_{\bar{k}}^{\infty}(\bar{k}-\frac{1}{2})\bar
{\varepsilon}_{\bar{k}}}
\nonumber\\
&=&q^{-\frac{c}{24}+\frac{1}{16}}\bar q^{-\frac{c}{24}}\prod_{k=1}^\infty \left(1+q^k\right)
\prod_{\bar k=1}^\infty\left(1+\bar q^{\bar k -\frac{1}{2}} \right ).
\nonumber
\end{eqnarray}
Taking into account the expressions for the chiral characters $\chi_0, \chi _{\frac{1}{2}}, \chi _{\frac{1}{16}}$ \cite{henkel} and adding the conjugate part of the eigenvalue set, we can get the universal conformal partition function given by  Eq. (\ref{Z12}). Thus we have obtained analytically the universal conformal partition function for the Ising model with duality twisted BC's $Z_{(1,2)}(q)$ which confirms the numerical result of \cite{Pearce2001}.

Now we will present the new set of the universal amplitude ratios. Let us denote the free energy per spin $f$, the inverse correlation lengths $\xi^{-1}_{p}$ and $\xi^{-1}_{L-p}$ of our critical Ising model as $L_x f=-\frac{\sqrt{2}}{2}\log\left(\left|\Lambda_0\right|\right)$,  $\xi^{-1}_{p}=\frac{\sqrt{2}}{2}\log\left(\left|\frac{\Lambda_0}{\Lambda_{\{p\}}}\right|\right)$ and
$\xi^{-1}_{L-p}=\frac{\sqrt{2}}{2}\log\left(\left|\frac{\Lambda_0}{\Lambda_{\{L-p\}}}\right|\right)$. We find that subdominant finite-size corrections to scaling should be to the form $a_k/L^{2k-1}_x$ for the free energy $f$ and $b_k(p)/L_x^{2k-1}$ and $c_k(p)/L_x^{2k-1}$ for inverse correlation lengths $\xi^{-1}_p$ and $\xi^{-1}_{L-p}$, respectively, where the coefficients $a_k, b_k(p)$ and $c_k(p)$ can be written in the following form
\begin{eqnarray}
a_k&=&\frac{\pi^{2k-1}B_{2k}}{2^{k-1}(2k)!}f^{(2k-1)}(0)
\nonumber\\
b_{k}(p)&=&\frac{\pi^{2k-1}(2p-1)^{2k-1}}{2^{k-2}(2k-1)!}f^{(2k-1)}(0)
\nonumber\\
c_{k}(p)&=&\frac{2^{k+1}\pi^{2k-1}p^{2k-1}}{(2k-1)!}f^{(2k-1)}(0).
\nonumber
\end{eqnarray}
The coefficients $a_1, b_1(p)$ and $c_1(p)$ are universal and related  to the conformal anomaly number ($c$), the conformal weights ($\Delta, \bar \Delta$), and the scaling dimensions
of the p-th scaling fields of the theory. The coefficients $a_k$, $b_{k}(p)$ and $c_{k}(p)$ for $k \ge 2$ are non-universal, but ratios of these coefficients $r_p(k)=\frac{b_{k}(p)}{a_k}$ and $r_{L-p}(k)=\frac{c_{k}(p)}{a_{k}(p)}$ are universal and given by
\begin{eqnarray}
r_p(k)=\frac{4 k(2p-1)^{2k-1}}{B_{2k}}; \quad r_{L-p}(k)=\frac{4 k(2p)^{2k-1}}{B_{2k}}.
\nonumber
\end{eqnarray}
Thus we have obtained a new set of the universal amplitude ratios $r_p(k)$ and $r_{L-p}(k)$. The case for $k=1$ is trivial, since $r_p(1), r_{L-p}(1)$ are the ratios of universal coefficients $a_1$, $b_{1}(p)$ and $c_{1}(p)$. The case $k \geq 2$ is non-trivial. Below we will show that the ratios $r_p(2)$ and $r_{L-p}(2)$ which are given by
\begin{eqnarray}
r_p(2) &=& -240 (2p-1)^3\label{r2}\\
r_{L-p}(2)&=& -1920 p^3 \label{rL2}
\end{eqnarray}
can be obtain from conformal field theory.
The finite-size corrections to Eq.~(\ref{energy}) can in principle be computed
in perturbative conformal field theory. In general, any critical lattice Hamiltonian
will contain correction terms to the fixed-point Hamiltonian $H =\zeta H_c + \sum_k g_k \int_{-L_x/2}^{L_x/2}\phi_k(v) d v$,
where $g_k$ is a non-universal constant and $\phi_k(v)$ is a
perturbative conformal field with scaling dimension $x_k$. To the first order in the perturbation, the energy gaps $({\cal E}_n-{\cal E}_0)$ and the ground-state energy (${\cal E}_0$) can be
written as
\begin{eqnarray}
{\cal E}_n-{\cal E}_0&=&\frac{2 \pi}{L_x}\zeta x_n+ 2 \pi \sum_k
g_k(C_{nkn}-C_{0k0})\left(\frac{2 \pi}{L_x}\right)^{x_k-1},
\nonumber\\
{\cal E}_0 &=& {\cal E}_{0,c}+2 \pi \sum_k g_k C_{0k0} \left(\frac{2
\pi}{L_x}\right)^{x_k-1}, \nonumber
\end{eqnarray}
where $C_{nkn}$ are universal structure constants. In the case of the cylinder the spectra of the Hamiltonian  are built by
the irreducible representation $\Delta, \bar \Delta$ of two
commuting Virasoro algebras $L_n$ and ${\bar L}_n$.
The leading finite-size corrections ($1/L_x^3$) can be described by the Hamiltonian  with a single perturbative conformal field
$\phi_1(v)=L_{-2}^2(v)+\bar L_{-2}^2(v)$ with scaling dimension $x_1=4$ \cite{henkel}. Thus the ratio $r_n(2)$ are indeed universal and given by
\begin{equation}
r_n(2)=\frac{C_{n1n}-C_{010}}{C_{010}}.
\nonumber
\end{equation}
The universal structure constants $C_{n1n}$ can be obtained from the matrix elements $\langle n|\phi_1(0)|n \rangle =\left({2 \pi}/{L_x}\right)^{x_1}C_{n1n}$ \cite{cardy86}, which for non-degenerated states have already been computed by Reinicke \cite{reinicke87}:
\begin{eqnarray}
C_{n1n}&=& (\Delta+r)\left(\Delta-\frac{2+c}{12}+\frac{r(2 \Delta +
r)(5 \Delta+1)}{(\Delta+1)(2\Delta+1)}\right)
\label{cnln}\\
&+&\left(\frac{c}{24}\right)^2+\frac{11 c}{1440} +\frac{r}{30}
 \left[r^2(5c-8)-(5c+28)\right]\delta_{\Delta,0}. \nonumber
\end{eqnarray}
Let us consider the case $c=1/2$. For the two-dimensional Ising model with duality twisted BC's the ground state $|0>$, the excited state $|p>$ and the excited state $|L-p>$ are given by $|0>=|\Delta_0=\frac{1}{16},r=0;\bar \Delta_0=0,\bar r=0>$, $|p>=|\Delta_p=\frac{1}{16},r=0;\bar \Delta_p=\frac{1}{2},\bar r=p-1>$ and $|L-p>=|\Delta_{L-p}=\frac{1}{16},r=p;\bar \Delta_{L-p}=0,\bar r=0>$.

For non-degenerated states the universal structure constants $C_{n1n}$ ($n=0, p$ and $L-p$) can be obtained from Eq. (\ref{cnln}) and for ratios $r_p(2)$ and $r_{L-p}(2)$ one can obtain
\begin{eqnarray}
r_p(2)&=&
-240(2p-1)^3\label{rpconf}\\
r_{L-p}(2)&=&
-\frac{1920}{17}p(16p^2+3p-2).\label{rLpconf}
\end{eqnarray}
Thus we can see that Eq. (\ref{rpconf}) coincides with Eq. (\ref{r2}) for all values of $p$, while Eq. (\ref{rLpconf}) coincides with Eq. (\ref{rL2}) only for $p=1$ and $2$. The excited states $|L-p>$ for $p \ge 3$ are degenerated and one cannot apply the Reinicke formula (\ref{cnln}). For degenerated states the calculations of the universal structure constants $C_{nkn}$ is not straightforward, but for the case of the Ising model it can be done. For the Ising model we have calculated the universal structure constants $C_{nkn}$ for degenerated states and find that the results are in complete agreement with Eq. (\ref{rL2}) for all values of $p$.

In this Letter we considered the Ising model with  duality twisted BC's.
We have showen that BC's with $(\Delta, \bar \Delta)= \left(\frac{1}{16}, 0\right)$  correspond to the seam $(r,s)=(1,2)$. We derive exact expressions for ${\it all}$ eigenvalues of the transfer matrix for the critical Ising model with the duality twisted BC's.
We derive explicitly the universal conformal partition function $Z_{(1,2)}(q)$ and show that it is consistent with numerical results of \cite{Pearce2001}. We find that the ratio of the subdominant finite-size corrections to scaling in the asymptotic expansion of the free energy $f$ and the inverse correlation lengths $\xi_p^{-1}$ and $\xi_{L-p}^{-1}$  are universal and give their exact value. We verify this universal behavior using a perturbative conformal approach.

{\bf Acknowledgments:} We would like to thank Paul Pearce and Rubik Poghossian for discussions.
This research was supported by  Marie Curie  grants (SPIDER, PIRSES-GA-2011-295302; RAVEN, MC-IIF-300206; DIONICOS, PIRSES-GA-2013-612707) and by a grant of the  Ministry of Science and Education of the Republic of Armenia (contract 13-1C080).

\end{document}